\documentclass[a4paper,amsmath,showpacs,showkeys,aps,nofootinbib,reprint]{revtex4-1}

\begin{document}

\title{Top quark as a resonance}
\author{A.E. Kaloshin}
\email{\url{kaloshin@physdep.isu.ru}}
\affiliation{Irkutsk State University, Irkutsk, Russia 664003}

\author{V.P. Lomov}
\email{\url{lomov.vl@icc.ru}, \url{V.P.Lomov@gmail.com}}
\affiliation{Institute for System Dynamics and Control Theory of SB RAS, Irkutsk, Russia 664033}
\affiliation{Irkutsk State Technical University, Irkutsk, Russia 664074}

\begin{abstract}
  We suggest the description of the dressed fermion propagator with parity non-conservation in
  the form with separated positive and negative energy poles. We found general form of the
  $\gamma$-matrix off-shell projectors and corresponding resonance factors. The parity violation
  leads to deviation of resonance factors from the naive Breit--Wigner form and to appearance of
  non-trivial spin corrections. However, for top quark with SM vertex the resonance factor
  returns to the standard one due to $\Gamma/m\ll1$.
\end{abstract}

%%% PACS numbers, the full scheme can be found on address http://www.aip.org/pacs/
%%% Used version: 2010 ed.
%%% Extract from PACS:
%%% 12.39.Hg Heavy quark effective theory
%%% 12.39.Ki Relativistic quark model
%%% 14.65.Ha Top quarks

\maketitle

\section{Introduction}
\label{sec:introduction}

The top quark, the heaviest elementary particle observed to date, plays a special role in
Standard Model (SM) \cite{Wagner:2005jh,Quadt:2007jk,Incandela:2009pf} and it is an object of
intensive research at LHC
\cite{Khachatryan:2010ez,Chatrchyan:2011nb,Aad:2010ey,Aad:2011yb,Bernreuther:2008ju}. %,Wicke:2010cg,Han:2008xb}.
Being a short-living particle (due to the open channels with $\pW$-boson on mass shell), it may
be considered on an equal footing with ordinary hadron resonances. The dressed propagator can be
obtained as a result of Dyson summation of self-energy insertions or, equivalently, by solving
the Dyson--Schwinger equation. As for top quark, its vertex violates parity, so $\gamma^{5}$
takes part in this process, and it leads to nonstandard form of resonance factor, as we shall
see below.

The form of fermion resonance with parity violation was discussed earlier. In particular, in
\cite{Kaloshin:2004jh} were written general formulas for dressed propagator with the use of the
off-shell basis. The paper \cite{Kniehl:2008cj} was devoted to extension of the concept of pole
mass and width \cite{Sirlin:1991fd,*Sirlin:1991rt,Sirlin:1998ps,Gambino:1999ai,Nekrasov:2001nh}
to the case of the parity violation. The obtained dressed fermion propagator was written in a
boson-like form without separation of the positive and negative energy poles. It is difficult to
compare this general expression with the standard Breit--Wigner form, in particular to recognize
there the on-shell decay width.

In this work we make the next step: we build the $\gamma$-matrix projectors onto the positive
and negative energy poles and corresponding resonance factors (see
\eqref{eq:DS-props-ferm-case-projs} and its generalization \eqref{eq:fp-projs-decomp}). The key
moment is the use of the spectral representation of operator for this purpose. The explicit form
of this representation \eqref{eq:ip-projs}, \eqref{eq:fp-projs-decomp} can be obtained for
arbitrary form of interaction, its particular case \eqref{eq:propag-in-vicinity} corresponds to
V-A vertex of SM.

On the way we discuss general relation \eqref{eq:width-relation} between on-shell width and
imaginary part of self-energy component, which is valid for any form of interaction and for
parity violation in particular.

\section{Standard Breit--Wigner formula in QFT}
\label{sec:stand-breit-wign}

To obtain Breit--Wigner-like formula in Quantum Field Theory (QFT) one needs to solve the
Dyson--Schwinger equation for the dressed propagator,
\begin{equation}
  \label{eq:DS-gf}
  G=G_{0}+G_{0}\Sigma G, \quad\text{or}\quad G^{-1}=G_{0}^{-1}-\Sigma,
\end{equation}
where $G$ and $G_{0}$ are dressed and free propagators and $\Sigma$ is a self-energy.

For bosons one has
\begin{equation*}
  G_{0}=\frac{1}{m_{0}^{2}-s-\imath\varepsilon}
\end{equation*}
and equation \eqref{eq:DS-gf} gives
\begin{equation}
  \label{eq:DS-props-bos-case}
  G=\frac{1}{m_{0}^{2}-s-\Sigma(s)}\sim\frac{1}{m^{2}-s-\imath\Gamma m}
\end{equation}
and, if $\Sigma$ has imaginary part, the dressed propagator $G$ should be compared with
relativistic Breit--Wigner formula for renormalization.

For fermion propagator one has
\begin{equation}
  \label{eq:DS-props-ferm-case}
  G_{0}=\frac{1}{\hat{p}-m_{0}}\quad\text{and}\quad
  G=\frac{1}{\hat{p}-m_{0}-\Sigma(p)},
\end{equation}
but to make this procedure more transparent, it is convenient to pass to off-shell projection
operators.

Let us define off-shell projection operators as follows:
\begin{equation}
  \label{eq:os-projs}
  \Lambda^{\pm}=\frac{1}{2}\Big(1\pm\frac{\hat{p}}{W}\Big),
\end{equation}
where $W=\sqrt{p^{2}}$ is invariant mass or rest-frame energy.

In this basis $G_{0}$ is
\begin{equation*}
  G_{0} = \frac{1}{\hat{p}-m_{0}}=\Lambda^{+}\frac{1}{W-m_{0}}+\Lambda^{-}\frac{1}{-W-m_{0}}
\end{equation*}
and solution of Dyson--Schwinger equation looks like\footnote{From our point of view, this
  representation for dressed fermion propagator is physically justified. If we are concerned
  with baryon resonance production, $\pi N\to N'\to\pi N$, then coefficients at $\Lambda^{\pm}$
  in \eqref{eq:DS-props-ferm-case-projs} appear in different partial waves and it makes no sense
  to join them together. Besides, it has long been known that proper variable for fermions is
  $W$, not $s$, see e.g. MacDowell symmetry \cite{MacDowell:1959zza}, when $W\to-W$.}
\begin{equation}
  \label{eq:DS-props-ferm-case-projs}
  G = \Lambda^{+}\frac{1}{W-m_{0}-\Sigma_{1}(W)}+\Lambda^{-}\frac{1}{-W-m_{0}-\Sigma_{2}(W)},
\end{equation}
where the self-energy is also decomposed in this basis
\begin{multline}
  \Sigma(p)=A(p^{2})+\hat{p}B(p^{2})=\Lambda^{+}(A+WB)+\Lambda^{-}(A-WB)\equiv\\
  \equiv\Lambda^{+}\Sigma_{1}(W)+\Lambda^{-}\Sigma_{2}(W).
\end{multline}

The positive energy pole should be compared with Breit--Wigner formula
\begin{equation}
  \label{eq:BW-form-ferm-case}
  \frac{1}{W-m_{0}-\Sigma_{1}(W)}\sim \frac{1}{W-m+\imath\Gamma/2}.
\end{equation}

The above formulas correspond to the parity conservation, because the self-energy does not
involve $\gamma^{5}$.

\section{Dressed fermion propagator with parity violation}
\label{sec:ferm-reson-with}

In case of parity violation the projection basis \eqref{eq:os-projs} must be supplemented by
elements with $\gamma^{5}$, it is handy to choose the basis as \cite{Kaloshin:2004jh}
\begin{equation}
  \label{eq:os-b-projs}
  \operP_{1}=\Lambda^{+},\quad\operP_{2}=\Lambda^{-},\quad
  \operP_{3}=\Lambda^{+}\gamma^{5},\quad\operP_{4}=\Lambda^{+}\gamma^{5}.
\end{equation}
Now the decomposition of a self-energy or a propagator has four terms
\begin{equation}
  \label{eq:s-os-decomp}
  S=\sum_{M=1}^{4}S_{M}\operP_{M},
\end{equation}
where coefficients $S_{M}$ are followed by obvious symmetry properties
\begin{equation}
  \label{eq:os-coeffs-props}
  S_{2}(W)=S_{1}(-W),\quad S_{4}(W)=S_{3}(-W)
\end{equation}
and are calculated as
\begin{equation}
  \label{eq:os-coeff-calc}
  \begin{aligned}
    S_{1} &= \frac{1}{2}\tr(\operP_{1}S), &\qquad  S_{2} &= \frac{1}{2}\tr(\operP_{2}S),\\
    S_{3} &= \frac{1}{2}\tr(\operP_{4}S), &\qquad  S_{4} &= \frac{1}{2}\tr(\operP_{3}S).
  \end{aligned}
\end{equation}

Let us denote by $S(p)$ and $S_{0}(p)$ the dressed and free inverse propagators,
respectively. With the use of decomposition \eqref{eq:s-os-decomp}, the Dyson--Schwinger
equation \eqref{eq:DS-gf} is reduced to the set of equations for scalar coefficients
\begin{equation}
  \label{eq:DS-scalar-coeffs}
  S_{M}=(S_{0})_{M}-\Sigma_{M},\quad M=1,\ldots,4.
\end{equation}
Considering the self-energy $\Sigma$ as a known value, we obtain the dressed propagator
\begin{equation}
  \label{eq:prop-os-decomp}
  G=\sum_{M=1}^{4}G_{M}\operP_{M},
\end{equation}
where the coefficients $G_{M}$ are
\begin{equation}
  \label{eq:prop-coeffs-explicit}
  G_{1}=\frac{S_{2}}{\Delta},\quad
  G_{2}=\frac{S_{1}}{\Delta},\quad
  G_{3}=-\frac{S_{3}}{\Delta},\quad
  G_{4}=-\frac{S_{4}}{\Delta},
\end{equation}
and $\Delta=S_{1}S_{2}-S_{3}S_{4}$.

In spite of simple answer \eqref{eq:prop-os-decomp}, it is inconvenient because the positive and
negative energy poles are not separated, compare with formula
\eqref{eq:DS-props-ferm-case-projs}. Note that in case of the parity non-conservation the
comparison with Breit--Wigner formula is not so evident.

\section{Spectral representation of propagator}
\label{sec:spectr-repr-prop}

In order to obtain the analog of representation \eqref{eq:DS-props-ferm-case-projs} in case of
parity violation, we need to separate the positive and negative energy poles contributions in a
dressed propagator. This problem is solved by the spectral representation of inverse
propagator\footnote{%
  In fact we used for our purposes the so called \emph{spectral representation} of operator
  (see, e.g. textbook \cite{Messiah:1961qm}). In quantum-mechanical notations it has the form:
\begin{equation*}
  \hat{A}=\sum_{i}\lambda_{i}\Pi_{i}=\sum_{i}\lambda_{i}\ket{i}\bra{i},
\end{equation*}
where $\ket{i}$ are eigenvectors
\begin{equation*}
  \hat{A}\ket{i}=\lambda_{i}\ket{i},
\end{equation*}
and $\Pi_{i}=\ket{i}\bra{i}$ are corresponding projectors.
}
\begin{equation}
  \label{eq:eprojs-decomp}
  S=\lambda_{1}\Pi_{1}+\lambda_{2}\Pi_{2},
\end{equation}
where $\Pi_{k}$ are projectors, satisfying the eigenstate problem
\begin{equation}
  \label{eq:eigen-project-problem}
  S\Pi_{k}=\lambda_{k}\Pi_{k}.
\end{equation}

The problem can be solved in the most general case. Let $S(p)$ is defined by decomposition
\eqref{eq:s-os-decomp} with arbitrary coefficients, the matrix $\Pi$ also can be written in such
form with coefficients $a_{M}$.

Substituting it into the eigenstate problem \eqref{eq:eigen-project-problem}, we find, after
some algebra, that $\lambda_{i}$ are roots of the equation
\begin{equation}
  \label{eq:ce-ev}
  \lambda^{2}-\lambda(S_{1}+S_{2})+(S_{1}S_{2}-S_{3}S_{4})=0,
\end{equation}
and solution of \eqref{eq:eigen-project-problem} is
\begin{equation}
  \Pi_{i}=\operP_{1}a_{1}^{i}+\operP_{2}a_{2}^{i}-\frac{S_{3}}{S_{1}-\lambda_{i}}a_{2}^{i}\operP_{3}-\frac{S_{4}}{S_{2}-\lambda_{i}}a_{1}^{i}\operP_{4}\label{eq:proj-ansatz}
\end{equation}
with arbitrary coefficients $a_{1}$, $a_{2}$.

In order \eqref{eq:proj-ansatz} to be a projector, $\Pi^{2}=\Pi$, we need only one additional
condition
\begin{equation}
  \label{eq:proj-coeff-cond}
  a_{2}=1-a_{1}.
\end{equation}
After it the orthogonality property $\Pi_{1}\Pi_{2}=\Pi_{2}\Pi_{1}=0$ defines $a_{1}$
coefficient
\begin{equation*}
  a_{1}^{1}=\frac{S_{2}-\lambda_{1}}{\lambda_{2}-\lambda_{1}},\quad
  a_{1}^{2}=-\frac{S_{2}-\lambda_{2}}{\lambda_{2}-\lambda_{1}}.
\end{equation*}
As result we have the projectors
\begin{equation}
  \label{eq:ip-projs}
  \begin{split}
    \Pi_{1} &=
    \frac{1}{\lambda_{2}-\lambda_{1}}\Bigl((S_{2}-\lambda_{1})\operP_{1}+(S_{1}-\lambda_{1})\operP_{2}-\\
    &\phantom{\frac{1}{\lambda_{2}-\lambda_{1}}\Bigl((S_{2}}
    -S_{3}\operP_{3}-S_{4}\operP_{4}\Bigr),\\
    \Pi_{2} &=
    \frac{1}{\lambda_{1}-\lambda_{2}}\Bigl((S_{2}-\lambda_{2})\operP_{1}+(S_{1}-\lambda_{2})\operP_{2}-\\
    &\phantom{\frac{1}{\lambda_{1}-\lambda_{2}}\Bigl((S_{2}}
    -S_{3}\operP_{3}-S_{4}\operP_{4}\Bigr).
  \end{split}
\end{equation}
with desired properties:
\begin{itemize}
\item $S\Pi_{k}=\lambda_{k}\Pi_{k}$, where an eigenvalue $\lambda_{k}$ is a root of equation
  \eqref{eq:ce-ev},
\item $\Pi_{k}^{2}=\Pi_{k}$,
\item  $\Pi_{1}\Pi_{2}=\Pi_{2}\Pi_{1}=0$,
\item $\Pi_{1}+\Pi_{2}=1$.
\end{itemize}
Note that in fact $S(p)$ has been rewritten in an equivalent form: using the explicit form of
projectors \eqref{eq:ip-projs} one sees that
\begin{equation}
  \label{eq:os-projs-ident}
  \sum_{M=1}^{4}\operP_{M}S_{M}\equiv\lambda_{1}\Pi_{1}+\lambda_{2}\Pi_{2}.
\end{equation}

The dressed propagator $G(p)$ is obtained by reversing of equation \eqref{eq:eprojs-decomp}
\begin{equation}
  \label{eq:fp-projs-decomp}
  G=\frac{1}{\lambda_{1}}\Pi_{1}+\frac{1}{\lambda_{2}}\Pi_{2}.
\end{equation}
The determinant $\Delta(W)$ of $S$ is
\begin{multline}
  \label{eq:ip-deter}
  \Delta(W)=S_{1}S_{2}-S_{3}S_{4}=\\
  =(W-m_{0}-\Sigma_{1})(-W-m_{0}-\Sigma_{2})-\Sigma_{3}\Sigma_{4},
\end{multline}
where $\Sigma_{i}(W)$ are self-energy components in our basis \eqref{eq:os-b-projs}. Free
propagator has poles at points $W=m_{0}$ and $W=-m_{0}$, the dressed one has them at $W=m$ and
$W=-m$. On the other hand, $\Delta(W)$ is equal to product of eigenvalues
\begin{equation}
  \label{eq:ip-deter-prop}
  \Delta(W)=\lambda_{1}(W)\lambda_{2}(W),
\end{equation}
so in the spectral representation of propagator \eqref{eq:fp-projs-decomp} the positive and
negative energy poles contributions are separated from each other. Therefore, the matrices
\eqref{eq:ip-projs} are projectors onto these poles.

Solving equation \eqref{eq:ce-ev}, we obtain eigenvalues at any form of self-energy:
\begin{multline}
  \label{eq:ev-4any-se}
  \lambda_{1,2}(W)=-\Big(m_{0}+\frac{\Sigma_{1}(W)+\Sigma_{2}(W)}{2}\Big)\pm\\
  \pm\sqrt{\Big(W-\frac{\Sigma_{1}(W)-\Sigma_{2}(W)}{2}\Big)^{2}+\Sigma_{3}\Sigma_{4}}.
\end{multline}

\section{Relation between decay width and self-energy}
\label{sec:relat-betw-decay}
For our purposes we will derive the known relation between decay width and imaginary part of
self-energy, taking into account our basis \eqref{eq:os-b-projs}. Below we consider the process
$f(p,M)\to f(q,m_{f})+V(k,m_{V})$ as a close example.

The on-shell decay width is defined as
\begin{equation}
  \label{eq:ddecay_width}
  \dd\Gamma=\frac{1}{2M}\abs{\mathcal{M}}^{2}\frac{\dd^{3}k}{(2\pi)^{3}2k_{0}}
    \frac{\dd^{3}q}{(2\pi)^{3}2q_{0}}(2\pi)^{4}\delta^{(4)}(p-q-k).
\end{equation}
Using the equality
\begin{equation*}
  \dd^{3}k=\dd^{4}k\cdot\delta(k^{2}-m_{V}^{2})2k_{0}\theta(k_{0}),
\end{equation*}
one can rewrite the width as a four-dimensional integral,
\begin{multline*}
  \Gamma=\frac{1}{2M}\int\frac{\dd^{4}k}{(2\pi)^{2}}\abs{\mathcal{M}}^{2}\cdot
  \delta(k^{2}-m_{V}^{2})\theta(k_{0})\times\\
  \times\delta\big((q-k)^{2}-m_{f}^{2}\big) \theta(q_{0}-k_{0}).
\end{multline*}
It looks like the discontinuity of a loop, calculated according to Landau--Cutkosky rule.

Let us write down the decay matrix element and corresponding self-energy. The matrix element is

{
  \centering
  \includegraphics{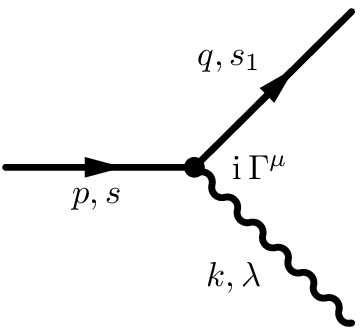}\par
}
\begin{equation}
  \label{eq:mat-el-vector}
  \mathcal{M}=\bar{u}(q) \Gamma^{\mu} u(p)\varepsilon_{\mu}(k),
\end{equation}
where the vertex $\Gamma_{\mu}$ contains the coupling constant and some $\gamma$-matrices.

The spinors are normalized as $\bar{u}u=-\bar{v}v=2m$. After summation and averaging over
polarizations we have
\begin{multline}
  \label{eq:matel-vec-case:sum-avr}
  \frac{1}{2}\sum_{s,s_{1},\lambda}\abs{\mathcal{M}}^{2}=
    -\frac{1}{2}\tr\Big((\hat{p}+M)\widetilde{\Gamma}_{\nu}
    (\hat{p}-\hat{k}+m_{f})\Gamma_{\mu}\Big)\times\\
    \times\Big(g^{\mu\nu}-\frac{k_{1}^{\mu}k_{1}^{\nu}}{m_{V}^{2}}\Big),\quad
    \text{where $\widetilde{\Gamma}_{\nu}=\gamma^{0}\Gamma^{\dag}_{\nu}\gamma^{0}$}.
\end{multline}
The width may be written as
\begin{multline}
  \label{eq:width-with-trace}
  \Gamma=-\frac{1}{2}\int\frac{\dd^{4}k}{(2\pi)^{2}}\tr\Big(\frac{\hat{p}+M}{2M}\widetilde{\Gamma}^{\nu}(\hat{p}-\hat{k}+m_{f})\Gamma^{\mu}\Big)\times\\
  \times\Big(g_{\mu\nu}-\frac{k_{\mu}k_{\nu}}{m_{V}^{2}}\Big)\cdot
  \delta(k^{2}-m_{V}^{2})\theta(k_{0})\cdot\delta((p-k)^{2}-m_{f}^{2})\theta(p^{0}-k^{0}).
\end{multline}

The corresponding self-energy is

{
  \centering
  \includegraphics{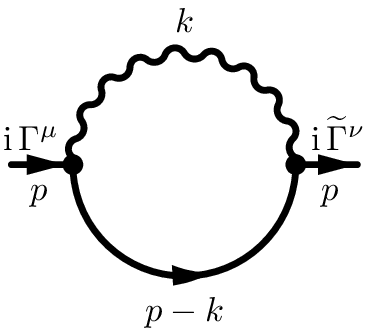}\par
}
\begin{equation}
  \label{eq:se-t-quark}
  \Sigma(p)=-\imath \int\frac{\dd^{4}k}{(2\pi)^{4}}\tilde{\Gamma}_{\nu}
    \frac{\hat{p}-\hat{k}+m_{f}}{(p-k)^{2}-m_{f}^{2}}
    \Gamma_{\mu}\frac{g^{\mu\nu}-k^{\mu}k^{\nu}/m_{V}^{2}}{k^{2}-m_{V}^{2}},
\end{equation}
and its discontinuity
\begin{multline}
  \label{eq:vs-se-disc}
  \Delta\Sigma(p)=\imath \int\frac{\dd^{4}k}{(2\pi)^{2}}
    \tilde{\Gamma}_{\mu}(\hat{p}-\hat{k}+m_{f}) \Gamma_{\nu} \big(g^{\mu\nu}-k^{\mu}k^{\nu}/m_{V}^{2}\big)\times\\
    \times\delta(k^{2}-m_{V}^{2})\theta(k_{0})\cdot\delta((p-k)^{2}-m_{f}^{2})\theta(p_{0}-k_{0}).
\end{multline}

The rules \eqref{eq:os-coeff-calc} allow to find out the coefficient $\Sigma_{1}$
\begin{equation*}
  \Sigma_{1}=\frac{1}{2}\tr(\operP_{1}\Sigma)=\frac{1}{2}\tr\Big(\frac{\hat{p}+W}{2W}\Sigma\Big).
\end{equation*}
Comparing it with the width \eqref{eq:width-with-trace} we obtain the following relation between
the width and self-energy in OMS
\begin{equation}
  \label{eq:width-relation}
  \Im{\Sigma_{1}(W=M)}=-\frac{\Gamma}{2}.
\end{equation}

Let us stress that the relationship \eqref{eq:width-relation} is very general, it is valid for
any final state in a decay. Moreover, as is seen from our derivation, it does not depend on
whether the parity is conserved or not.

\section{$\pt$-quark propagator in Standard Model}
\label{sec:standard-model}

Consider the dressing of top quark in SM. The main one-loop contribution to self-energy arises
from $\pW\pb$ intermediate state
\begin{multline}
  \label{eq:one-loop-cor}
  \Sigma(p)=-\imath g^{2}\abs{V_{\pt\pb}}^{2}\int\frac{\dd^{4}k}{(2\pi)^{4}}\gamma^{\mu}(1-\gamma^{5})
    \frac{\hat{p}-\hat{k}+m_{\pb}}{(p-k)^{2}-m^{2}_{\pb}}\times\\
    \times\gamma^{\nu}(1-\gamma^{5})\frac{g_{\mu\nu}-k_{\mu}k_{\nu}/m^{2}_{\pW}}{k^{2}-m^{2}_{\pW}},
\end{multline}
and generates only kinetic term
\begin{equation}
  \label{eq:t-quark-loop}
  \Sigma(p)=\hat{p}(1-\gamma^{5})\Sigma_{0}(W^{2}).
\end{equation}

Its decomposition in the basis \eqref{eq:os-b-projs} has the following coefficients:
\begin{multline}
  \label{eq:sigma-coeffs-4-t-quark}
  \Sigma_{1}=W\Sigma_{0}(W^{2}),\;
  \Sigma_{2}=-W\Sigma_{0},\\
  \Sigma_{3}=-W\Sigma_{0},\;
  \Sigma_{4}=W\Sigma_{0}.
\end{multline}

In this case, the general relation \eqref{eq:width-relation} gives
\begin{equation}
  \label{eq:t-quark-width-rel}
  \Im\Sigma_{0}(W^{2}=m^{2})=-\frac{\Gamma}{2m}.
\end{equation}

As a preliminary, let us forget about renormalization of self-energy and use
\eqref{eq:sigma-coeffs-4-t-quark} to calculate the eigenvalues\footnote{We wrote $\lambda_{i}$
  in a form respecting the symmetry property $\lambda_{2}(W)=\lambda_{1}(-W)$, valid for bare
  values.}
\begin{equation*}
  \lambda_{1,2}=-m\pm W\sqrt{1-2\Sigma_{0}(W^{2})}.
\end{equation*}
In analogy with on-mass-shell (OMS) scheme of renormalization, we should subtract the real part
of self-energy at resonance point
\begin{equation*}
  \lambda_{1,2}=-m\pm W\sqrt{1-2\big(\Sigma_{0}(W^{2})-\Re\Sigma_{0}(m^{2})\big)}.
\end{equation*}
As a result we have rather unusual resonance factor
\begin{equation}
  \label{eq:res-factor-naive}
  \frac{1}{\lambda_{1}(W)}=\frac{1}{W\sqrt{1+\imath\dfrac{\Gamma}{m}}-m},
\end{equation}
which only at $\Gamma/m\ll1$ returns to standard Breit--Wigner form,
\begin{equation*}
  \frac{1}{\lambda_{1}(W)}\simeq\frac{1}{W-m+\imath  W\dfrac{\Gamma}{2m}}
  \quad\text{at } \Gamma/m\ll1.
\end{equation*}

Of course, this nonstandard resonance factor arises due to parity violation: the appearance of
$\gamma^{5}$ in vertex leads to another Dyson summation of self-energy in a propagator.

To analyse the obtained dressed propagator in more detail, we need to renormalize it. We will
use the OMS scheme of renormalization in order to compare with Breit--Wigner formula.

At first, let us suppose that self-energy does not have imaginary part. The determinant
\eqref{eq:ip-deter} in case of parity violation resembles the similar object in the mixing
problem. This analogy allows to formulate the OMS requirements (see
\cite{Denner:1990ae,Aoki:1982ed}) on the self-energy
\begin{itemize}
\item $\Sigma_{1}$ has zero of second order at $W=m$
\item $\Sigma_{3}$ has zeroes at $W=m$ and $W=-m$.
\end{itemize}
The $\Sigma_{2}$ and $\Sigma_{4}$ are defined by substitution $W\to-W$, so the OMS
renormalization in this case is
\begin{flalign*}
  \Sigma^{\ren}_{1}(W) &= \Sigma_{1}(W)-\Sigma_{1}(m)-\Sigma'_{1}(m)(W-m),\\
  \Sigma^{\ren}_{2}(W) &= \Sigma^{\ren}_{1}(-W),\\
  \Sigma^{\ren}_{3}(W) &= -W\Big(\Sigma_{0}(W^{2})-\Sigma_{0}(m^{2})\Big),\\
  \Sigma^{\ren}_{4}(W) &= \Sigma^{\ren}_{3}(-W).
\end{flalign*}

Eigenvalues in OMS scheme are
\begin{equation}
  \label{eq:ev-oms-real-case}
  \lambda_{1,2}(W)=-mK\pm WK\sqrt{d},\;
  \text{where}\;
  d=1-2\widetilde{\Sigma}/K
\end{equation}
and $K=1+2m^{2}\Sigma'_{0}(m^{2})$, $\widetilde{\Sigma}=\Sigma_{0}(W^{2})-\Sigma_{0}(m^{2})$.

Let us write down the eigenvalues in vicinity of $W=m$
\begin{flalign*}
  \lambda_{1}(W) &= W-m+o(W-m),\\
  \lambda_{2}(W) &= -2mK-(W-m)+o(W-m),
\end{flalign*}
and in vicinity of $W=-m$
\begin{flalign*}
  \lambda_{1}(W) &= -2mK-(-W-m)+o(-W-m),\\
  \lambda_{2}(W) &= -W-m+o(-W-m).
\end{flalign*}

Projectors on eigenstates \eqref{eq:ip-projs} have the form\footnote{We want to pay attention on
  appearance of factor $K$ in these formulas, it arises because $\lambda_{1}$ and $\lambda_{2}$
  are normalized at different points: $W=m$ and $W=-m$ correspondingly.  Note that the natural
  variable for fermions is just $W$, but not $W^{2}$, it is well known, e.g. in
  $\pi\particle{N}$ scattering. This fact leads to some difference in resonance denominators of
  fermion and boson propagators, noted in \cite{Gonchar:2006xv}.}
\begin{equation}
  \label{eq:proj-vicin-v-a}
  \begin{split}
    \Pi_{1} &= \operP_{1}\frac{\sqrt{d}+(1-\widetilde{\Sigma}/K)}{2\sqrt{d}} +
    \operP_{2}\frac{\sqrt{d}-(1-\widetilde{\Sigma}/K)}{2\sqrt{d}}-\\
    &\phantom{\operP_{1}\frac{\sqrt{d}+(1-\widetilde{\Sigma}/K)}{2\sqrt{d}}} -
    \operP_{3}\frac{\widetilde{\Sigma}/K}{2\sqrt{d}}+
    \operP_{4}\frac{\widetilde{\Sigma}/K}{2\sqrt{d}},\\
    \Pi_{2} &= \operP_{1}\frac{\sqrt{d}-(1-\widetilde{\Sigma}/K)}{2\sqrt{d}}+
    \operP_{2}\frac{\sqrt{d}+(1-\widetilde{\Sigma}/K)}{2\sqrt{d}}+\\
    &\phantom{\operP_{1}\frac{\sqrt{d}-(1-\widetilde{\Sigma}/K)}{2\sqrt{d}}}+
    \operP_{3}\frac{\widetilde{\Sigma}/K}{2\sqrt{d}}-
    \operP_{4}\frac{\widetilde{\Sigma}/K}{2\sqrt{d}}.
  \end{split}
\end{equation}

Substituting formulas for eigenvalues \eqref{eq:ev-oms-real-case} and projectors
\eqref{eq:proj-vicin-v-a} back into \eqref{eq:fp-projs-decomp} we obtain
\begin{equation*}
  G(p) = \frac{m_{0}+\hat{p}-\hat{p}(1+\gamma^{5})\tSigma/K}{K(W^{2}d-m_{0}^{2})}.
\end{equation*}

The expressions for eigenvalues and projectors may be simplified in vicinity of $W^{2}=m^{2}$,
where $\widetilde{\Sigma}(W)\ll1$ and we take into account only linear in $\widetilde{\Sigma}$
terms
\begin{flalign*}
  \lambda_{1,2}(W) &= K(-m\pm W)\mp W\widetilde{\Sigma}(W^{2}),\\
  \Pi_{1} &= \operP_{1}-\operP_{3}\frac{\widetilde{\Sigma}}{2K}+\operP_{4}\frac{\widetilde{\Sigma}}{2K}=
  \Lambda^{+}-\frac{\widetilde{\Sigma}(W^{2})}{2K}\frac{\hat{p}\gamma^{5}}{W},\\
  \Pi_{2} &= \operP_{2}+\operP_{3}\frac{\widetilde{\Sigma}}{2K}-\operP_{4}\frac{\widetilde{\Sigma}}{2K}=
  \Lambda^{-}+\frac{\widetilde{\Sigma}(W^{2})}{2K}\frac{\hat{p}\gamma^{5}}{W}.
\end{flalign*}

Now let us return to the case when the self-energy $\Sigma(W)$ acquire the imaginary part. In
this situation we use a generalization \cite{Denner:1991kt,Bardin:1999ak} of OMS scheme for
unstable particles, which consists in subtraction of real part of a loop. The formulas for
eigenvalues and projectors, \eqref{eq:ev-oms-real-case} and \eqref{eq:proj-vicin-v-a}, remain
the same, but in this case
\begin{align*}
  \widetilde{\Sigma}(W^{2})&=\Sigma_{0}(W^{2})-\Re\Sigma_{0}(m^{2}),\quad\text{and}\\
  K&=1+2m^{2}(\Re\Sigma_{0})'(m^{2}).
\end{align*}
Resonance factor $1/\lambda_{1}$ in vicinity of $W=m$ practically coincides with naive
expression \eqref{eq:res-factor-naive}
\begin{multline}
  \label{eq:res-factor-complex}
  \frac{1}{\lambda_{1}(W)}=\frac{1}{K\Big(W\sqrt{1-2\widetilde{\Sigma}/K}-m\Big)}\approx\\
  \approx\frac{1}{K\Big(W\sqrt{\vphantom{\widetilde{\Sigma}}1+\imath\frac{\Gamma(W)}{KW}}-m\Big)},
\end{multline}
if to introduce the energy-dependent width $\Gamma(W)=-2W\Im\Sigma_{0}(W^{2})$.

At small $\Gamma$ resonance factor returns to standard form
\begin{equation*}
  \frac{1}{\lambda_{1}(W)}\simeq\frac{1}{W-m+\imath\Gamma(W)/2}\quad
  \text{at } W\simeq m,\,\Gamma/m\ll1.
\end{equation*}
Using the same approximations in projectors, we can write down a parametrization of dressed
propagator in vicinity of $W=m$:
\begin{multline}
  \label{eq:propag-in-vicinity}
  G=\frac{1}{W-m+\imath\Gamma(W)/2}\Big(\operP_{1}+\imath\frac{\Gamma(W)}{4KW^{2}}\hat{p}\gamma^{5}\Big)+\\
   +\frac{1}{-2mK-(W-m)-\imath\Gamma(W)/2}\Big(\operP_{2}-\imath\frac{\Gamma(W)}{4KW^{2}}\hat{p}\gamma^{5}\Big).
\end{multline}

Let us compare our expression \eqref{eq:propag-in-vicinity} for propagator with one given in
\cite{Kniehl:2008cj}:
\begin{equation}
  \label{eq:ks-propag}
  \imath S^{(r)}(p)=\imath\big[S_{+}^{(r)}(p)a_{+}+S_{-}^{(r)}(p)a_{-}\big],
\end{equation}
where $a_{\pm}=(1\pm\gamma^{5})/2$. Note that in \eqref{eq:ks-propag} both terms have positive
and negative poles contributions.

\section{Pole scheme and spectral representation}
\label{sec:pole-renorm-scheme}

The pole renormalization scheme for fermion with parity non-conservation have been considered in
detail in \cite{Kniehl:2008cj}. We will consider the pole scheme on the base of spectral
representation. Instead of renormalization of determinant, as in \cite{Kniehl:2008cj}, in this
case it is sufficient to renormalize the single pole contribution $1/\lambda_{1}(W)$. It
simplifies essentially the algebraic procedure and clarifies some aspects.

The inverse propagator has the form
\begin{multline}
  \label{eq:inv-prop}
  S(p)=\hat{p}-m_{0}-\Sigma(p)=\\
  =\hat{p}-m_{0}-\big(A(p^{2})+\hat{p}B(p^{2})+C(p^{2})\gamma^{5}+\hat{p}\gamma^{5}D(p^{2})\big).
\end{multline}

In $\mathsf{CP}$-symmetric theory $C(p^{2})=0$, see \cite{Kniehl:2008cj}.

In terms of scalar functions the eigenvalues and corresponding projectors \eqref{eq:ip-projs}
have the form
\begin{equation}
  \label{eq:eigens-explicit}
  \begin{split}
    \lambda_{1}(W) &= -m_{0}-A(W^{2})+W R(W^{2}),\\
    \lambda_{2}(W) &= \lambda_{1}(-W),
  \end{split}
\end{equation}
\begin{equation}
  \label{eq:proj1-pole}
  \begin{split}
    \Pi_{1}(W) &= \frac{1}{2}\Big[1-\gamma^{5}\frac{C(W^{2})}{W R(W^{2})}+\\
      &+\frac{\hat{p}}{W}\Big(\frac{1-B(W^{2})}{R(W^{2})}-\gamma^{5}\frac{D(W^{2})}{R(W^{2})}\Big)\Big],\\
    \Pi_{2} &= \Pi_{1}(-W),
  \end{split}
\end{equation}
where we have introduced the notation
\begin{equation}
  \label{eq:R-func}
  R(W^{2})=\sqrt{\big(1-B(W^{2})\big)^{2}-D^{2}(W^{2})+C^{2}(W^{2})/W^{2}}.
\end{equation}

Let's $\lambda_{1}(W_{1})=0$, where $W_{1}=M_{p}-\imath \Gamma_{p}/2$:
\begin{equation}
  \label{eq:ev-pole-zero}
  -m_{0}-A(W_{1}^{2})+W_{1} R(W_{1}^{2})=0.
\end{equation}

Real part of this equality allows to get rid of $m_{0}$ in dressed propagator
\begin{gather*}
  S(p)=\hat{p}-\Big(\tilde{A}(p^{2})+\hat{p}B(p^{2})+\gamma^{5}C(p^{2})+\hat{p}\gamma^{5}D(p^{2})\Big),\\
  \tilde{A}(p^{2})=A(p^{2})-A(W_{1}^{2})+\big(W_{1} R(W_{1}^{2})\big).
\end{gather*}

The imaginary part of \eqref{eq:ev-pole-zero},
\begin{equation}
  \label{eq:ev-zero-im-part}
  \Im\Big(-A(W_{1}^{2})+W_{1} R(W_{1}^{2})\Big)=0
\end{equation}
gives relation between $\Gamma_{p}$ and self-energy at pole point. In particular, in case of
parity conservation it reduces to the obvious relation
\begin{equation}
  \label{eq:im-se-obv}
  \begin{gathered}
    \Im\Big(W_{1}-\big(A(W_{1}^{2})+W_{1} B(W_{1}^{2})\big)\Big)=0,\\
    \text{or}\quad
    \frac{\Gamma_{p}}{2}=-\Im\Sigma_{1}(W_{1}^{2}).
  \end{gathered}
\end{equation}

It is possible to express $\Gamma_{p}$ from \eqref{eq:ev-pole-zero} in other way, as was done in
\cite{Kniehl:2008cj}, but then the expression will contain the bare mass $m_{0}$.

Let us introduce wave function renormalization constants connecting bare and renormalized fields
\begin{equation}
  \label{eq:mr-def}
  \Psi=Z^{1/2}\Psi^{\ren},\quad
  \bar{\Psi}=\bar{\Psi}^{\ren}\bar{Z}^{1/2}.
\end{equation}
In case of parity violation $Z^{1/2}$, $\bar{Z}^{1/2}$ are matrices\footnote{We prefer to use
  the form \eqref{eq:z-zbar-def} instead of chiral projectors $a_{\pm}=(1\pm\gamma^{5})/2$ to
  have more simple intermediate expressions.}
\begin{equation}
  \label{eq:z-zbar-def}
  Z^{1/2}=\alpha+\beta\gamma^{5},\quad
  \bar{Z}^{1/2}=\bar{\alpha}+\bar{\beta}\gamma^{5}.
\end{equation}
Relations between parameters are discussed below.

Renormalized inverse propagator
\begin{multline}
  \label{eq:inv-renorm-prop}
  S^{\ren}(p)=(\bar{\alpha}+\bar{\beta}\gamma^{5})\Big[\hat{p}-\big(\tilde{A}+\hat{p}B+
    \gamma^{5}C+\hat{p}\gamma^{5}D\big)\Big](\alpha+\beta\gamma^{5})=\\
    \begin{aligned}
      &= I\big[-\tilde{A}(\alpha\bar{\alpha}+\bar{\beta}\beta)-C(\bar{\alpha}\beta+\bar{\beta}\alpha)\big]+\\
      &+\hat{p}\big[(1-B)(\alpha\bar{\alpha}-\beta\bar{\beta})-D(\bar{\alpha}\beta-\bar{\beta}\alpha)\big]+\\
      &+\gamma^{5}\big[-C(\bar{\alpha}\alpha+\bar{\beta}\beta)-\tilde{A}(\bar{\alpha}\beta+\bar{\beta}\alpha)\big]+\\
      &+\hat{p}\gamma^{5}\big[-D(\bar{\alpha}\alpha-\bar{\beta}\beta)+(1-B)(\bar{\alpha}\beta-\bar{\beta}\alpha)\big]
    \end{aligned}
\end{multline}
allow to obtain the renormalized components of self-energy. Renormalized eigenvalues and
projectors are expressed through $A^{\ren}$, $B^{\ren}$, $C^{\ren}$ and $D^{\ren}$ by the same
formulas \eqref{eq:eigens-explicit} and \eqref{eq:proj1-pole}.

In our representation \eqref{eq:fp-projs-decomp} it is sufficient to renormalize only one pole
contribution, the second pole will obtain the correct properties automatically by substitution
$W\to-W$. Looking at first term in \eqref{eq:fp-projs-decomp}, we see that renormalization is
divided into two parts: renormalization of eigenvalue and projector. It is convenient to start
from renormalized projector $\Pi_{1}^{\ren}$.

For stable fermion there is a physical requirement for projector. As is seen from
\eqref{eq:proj1-pole} the projector at point $W=m$ has form
\begin{equation*}
  \Pi^{\ren}_{1}(m)=\frac{1}{2}\Big[1-\gamma^{5}c+\frac{\hat{p}}{m}\big(b-\gamma^{5}d\big)\Big],
\end{equation*}
where parameters $b$, $d$ and $c$ are related by $b^{2}-d^{2}+c^{2}=1$.  However, if $c\neq0$,
$d\neq0$ then $\Pi^{\ren}_{1}(m)$ do not commutate with spin projector, what leads to spin flip
for fermion on mass shell. Therefore there are requirements for renormalization of a stable
fermion:
\begin{equation}
  \label{eq:renorm-conds-stab-fer}
  C^{\ren}(m^{2})=0,\quad
  D^{\ren}(m^{2})=0.
\end{equation}

For unstable fermion, when pole is at point $W_{1}=M_{p}-\imath\Gamma_{p}/2$, there is some
arbitrariness. The simplest generalization of \eqref{eq:renorm-conds-stab-fer} consists in:
\begin{equation}
  \label{eq:ks-renorm-c-d}
  C^{\ren}(W_{1}^{2})=0,\quad
  D^{\ren}(W_{1}^{2})=0.
\end{equation}
The same relations arise from a principle, suggested in \cite{Kniehl:2008cj}: the chiral
components should have poles with unit absolute value of residue.

To see that, let us project the $\lambda_{1}$ pole contribution onto chiral components
\begin{equation}
  \label{eq:prop-chril-proj}
  G^{\ren}_{1}=\frac{1}{\lambda^{\ren}_{1}(W)}\Pi^{\ren}_{1}(a_{+}+a_{-})=G_{+}^{\ren}a_{+}+G_{-}^{\ren}a_{-},
\end{equation}
where
\begin{equation}
  \label{eq:prop-chril-comp}
  G_{\pm}^{\ren}=\frac{1}{\lambda^{\ren}_{1}(W)}\frac{1}{2}\Big[1\mp\frac{C^{\ren}}{WR^{\ren}}+
    \frac{\hat{p}}{W}\frac{1-B^{\ren}\mp D^{\ren}}{R^{\ren}}\Big].
\end{equation}
If to require the chiral components $G_{\pm}^{\ren}$ to have the projector of form
$(1+\hat{p}/W_{1})/2$ at resonance point, the conditions \eqref{eq:ks-renorm-c-d} are necessary.

A few words about the relation between renormalization constants $Z^{1/2}$, $\bar{Z}^{1/2}$. The
pseudo-hermiticity condition
\begin{equation}
  \label{eq:mr-ph}
  \bar{Z}^{1/2}=\gamma^{0}\big(Z^{1/2}\big)^{\dag}\gamma^{0},
\end{equation}
is traditionally used in literature, which is reduced to $\bar{\alpha}=\alpha^{*}$,
$\bar{\beta}=-\beta^{*}$. However, as it was noted in \cite{Espriu:2002xv}, one should refused
from this condition, if self-energy has absorptive parts. The same is seen from our renormalized
propagator \eqref{eq:inv-renorm-prop}. Assuming pseudo-hermiticity we calculate
$D^{\ren}(W^{2})$ thus:
\begin{multline}
  \label{eq:mr-ph-d}
  D^{\ren}(W^{2})=\abs{\alpha}^{2}\Big\{D(W^{2})\Big(1+\frac{\abs{\beta}^{2}}{\abs{\alpha}^{2}}\Big)-\\
    -\big(1-B(W^{2})\big)\Big(\frac{\beta}{\alpha}+\frac{\beta^{*}}{\alpha^{*}}\Big)\Big\}.
\end{multline}
Because $D(W^{2})$ and $B(W^{2})$ contain physically different contributions we cannot provide
the condition $D^{\ren}(W_{1}^{2})=0$ for complex self-energy. So, the pseudo-hermiticity
condition, tacitly assumed in \cite{Kniehl:2008cj}, seems to be too restrictive for parity
violating theory.

Let us consider below the case of $\mathsf{CP}$ conservative theory when component
$C(p^{2})=0$. In order to avoid $\mathsf{CP}$ violation under renormalization it is necessary to
require (see \eqref{eq:inv-renorm-prop})
\begin{equation}
  \label{eq:cp-conser-cond}
  \bar{\alpha}\beta+\bar{\beta}\alpha=0.
\end{equation}

The pseudo-hermiticity condition \eqref{eq:mr-ph} leads to \eqref{eq:cp-conser-cond} in case of
real $\alpha$, $\beta$ (stable fermion). However, for resonance one have to refuse from
\eqref{eq:mr-ph}.

Let us require condition \eqref{eq:cp-conser-cond} and express $\bar{\beta}$ from it
\begin{equation}
  \label{eq:beta-bar}
  \bar{\beta}=-\bar{\alpha}\frac{\beta}{\alpha}.
\end{equation}
Then renormalized inverse propagator becomes
\begin{equation}
  \label{eq:inv-prop-cp-cons-ren}
  \begin{split}
    S^{\ren} &= \alpha\bar{\alpha}\Big\{-\tilde{A}(W^{2})(1-x^{2})+\\
     &+\hat{p}\Big[(1-B(W^{2}))(1+x^{2})-D(W^{2})2x\Big]+\\
     &+\hat{p}\gamma^{5}\Big[-D(W^{2})(1+x^{2})+(1-B(W^{2}))2x\Big]\Big\},
  \end{split}
\end{equation}
where $\alpha$, $\bar{\alpha}$ and $x=\beta/\alpha$ are complex numbers.

The condition at pole $D^{\ren}(W_{1}^{2})=0$ defines\footnote{The requirement
  $\lambda_{1}^{\ren}(W_{1})=0$ allows unambiguously to fix $x$.}
\begin{equation}
  \label{eq:mr-x-factor}
  x\equiv\frac{\beta}{\alpha}=\frac{1-B_{1}-R_{1}}{D_{1}},
\end{equation}
where $B_{1}=B(W_{1}^{2})$, $D_{1}=D(W_{1}^{2})$, $R_{1}=R(W_{1}^{2})$. Substituting that into
$S^{\ren}$, taking out common factor and denoting it by $Z$ we get
\begin{equation}
  \label{eq:renorm-final}
  \begin{split}
    S^{\ren} &= Z\Big\{-\tilde{A}(W^{2})+\\
      &+ \hat{p}\Big[(1-B(W^{2}))\frac{1-B_{1}}{R_{1}}-D(W^{2})\frac{D_{1}}{R_{1}}\Big]+\\
      &+
      \hat{p}\gamma^{5}\Big[-D(W^{2})\frac{1-B_{1}}{R_{1}}+(1-B(W^{2}))\frac{D_{1}}{R_{1}}\Big]\Big\}=\\
      &=\hat{p}-\Sigma^{\ren},
  \end{split}
\end{equation}
where renormalized components are given by
\begin{align*}
  \tilde{A}^{\ren}(W^{2}) &= Z\tilde{A}(W^{2}),\\
  B^{\ren}(W^{2}) &= 1-Z\Big[\big(1-B(W^{2})\big)\frac{1-B_{1}}{R_{1}}-D(W^{2})\frac{D_{1}}{R_{1}}\Big],\\
  D^{\ren}(W^{2}) &= Z\Big[D(W^{2})\frac{1-B_{1}}{R_{1}}-\big(1-B(W^{2})\big)\frac{D_{1}}{R_{1}}\Big].
\end{align*}
To determine $Z$ factor let us consider renormalized eigenvalue $\lambda_{1}^{\ren}(W)$, its
derivative at $W=W_{1}$ has to equal 1. It is easy to check that
\begin{equation*}
  R^{\ren}(W^{2})=\sqrt{(1-B^{\ren}(W^{2}))^{2}-(D^{\ren}(W^{2}))^{2}}=Z R(W),
\end{equation*}
and
\begin{equation*}
  \lambda_{1}^{\ren}(W)=Z \lambda_{1}(W).
\end{equation*}
If to require $(\lambda_{1}^{\ren})'(W_{1})=1$ it gives
\begin{equation}
  \label{eq:mr-z-renorm-const}
  Z=\frac{1}{R(W_{1}^{2})+2W_{1}^{2}R'(W_{1}^{2})-2W_{1}A'(W_{1}^{2})}.
\end{equation}

% Finally, taking into account the expression for $Z$, we obtain following formulae for
% renormalized components
% \begin{align}
%   &
%   \begin{multlined}
%     A^{\ren}(W^{2}) = Z\tilde{A}(W^{2}) =\\
%     = Z\Big[A(W^{2})-A(W_{1}^{2})+W_{1} R(W_{1}^{2})\Big],
%   \end{multlined}\\
%   &
%   \begin{aligned}
%     B^{\ren}(W^{2}) &= Z\Big\{-2W_{1}A'(W_{1}^{2})+2W_{1}^{2}R'(W_{1}^{2})+\\
%                    &+ \big(B(W^{2})-B(W_{1}^{2})\big)\frac{1-B(W_{1}^{2})}{R(W_{1}^{2})}+\\
%                    &+ \big(D(W^{2})-D(W_{1}^{2})\big) \frac{D(W_{1}^{2})}{R(W_{1}^{2})}\Big\},
%   \end{aligned}\\
%   &
%   \begin{aligned}
%     D^{\ren}(W^{2}) &= Z\Big\{\big(D(W^{2})-D(W_{1}^{2})\big)\frac{1-B(W_{1}^{2})}{R(W_{1}^{2})}+\\
%                    &+ \big(B(W^{2})-B(W_{1}^{2})\big)\frac{D(W_{1}^{2})}{R(W_{1}^{2})}\Big\}.
%   \end{aligned}
% \end{align}

In case of unstable fermions, the right hand side of \eqref{eq:mr-z-renorm-const} is, generally
speaking, complex. If we define
\begin{equation}
  \label{eq:mr-ev-corr-def}
  \lambda_{1,2}^{\ren}(W)=\abs{Z}\lambda_{1,2}(W),
\end{equation}
we have the renormalized propagator with $\lambda_{i}(W)$ satisfying the Schwartz principle,
\begin{equation}
  \label{eq:ev-schratz-pr}
  \lambda_{i}^{\ren}(W^{*})=\big(\lambda_{i}^{\ren}(W)\big)^{*}.
\end{equation}
So, $\lambda_{i}^{\ren}$ has zeroes at complex conjugate points $W_{1}$, $W_{1}^{*}$ with unit
absolute value of residues.

\section{Conclusions}
\label{sec:conclusions}

We studied in detail the dressing of fermion propagator in the case of the parity
non-conservation. In contrast to previous works, we found the representation of propagator
\eqref{eq:ip-projs}, \eqref{eq:propag-in-vicinity}, where the positive and negative energy poles
are separated from each other. We compared our resonance representation with Breit--Wigner form
and used the on-shell definitions of mass and width. The spectral representation also allows to
perform pole renormalization in a simple and compact way.

We found that in case of parity violation the resonance factor \eqref{eq:res-factor-complex}
differs from Breit--Wigner-like formula. The reason is that in presence of $\gamma^{5}$ the
Dyson summation of the self-energy insertions in a propagator takes another form. But in case of
SM vertex the self-energy contains only the kinetic term and the obtained resonance factor
$1/\lambda_{1}(W)$ returns to the standard form for small width $\Gamma/m\ll1$.

For top quark $\Gamma/m\sim10^{-2}$ is really a small parameter, so for SM its resonance
propagator will practically coincide with standard one. Recall that at LHC the measurement of
$\Gamma_{\pt}$ is a rather challenging problem, and it is difficult to observe the deviation
from a standard picture in the form of a resonance curve.

Another possibility to see such a deviation is related with projectors \eqref{eq:ip-projs}. One
sees that $\Pi_{k}$ do not commutate with spin projectors $(1\pm\gamma^{5}\hat{s})/2$ and this
fact can lead to non-trivial spin properties at the level of $\Gamma/m$.

We suppose that spectral representation \eqref{eq:fp-projs-decomp} may be useful for neutrino
propagation and mixing, if to consider it in QFT approach. If, following
\cite{Grimus:1997aa,*Grimus:1998uh}, we are to consider neutrino propagation as macroscopic
Feynman diagram, then at long distances only the positive energy contribution survives, and
representation \eqref{eq:fp-projs-decomp} allows one to identify covariantly this term.

It is possible to generalize the spectral representation for matrix case, when the coefficients
in \eqref{eq:s-os-decomp} are matrices, and to use it for mixing problem with parity violation.

\section*{Acknowledgments}

We thank Slava Lee for participation at initial stage of the work and V.A. Naumov for discussion
and valuable advice.

\bibliographystyle{apsrev4-1} %%% aipnum4-1, apsrmp4-1
\bibliography{biblio3}

\end{document}